\documentstyle[10pt,psfig]{article}
\begin{document}
\begin{center} {\bf BLIND ITERATIVE DECONVOLUTION OF BINARY STAR IMAGES}
\end{center} 
\vspace{0.4cm}

\noindent
S.K. Saha and P. Venkatakrishnan \\
Indian Institute of Astrophysics, Bangalore 560034 \\
\vspace{0.4cm}

\noindent
{\bf Abstract} The technique of Blind Iterative De-convolution (BID) was used
to remove the atmospherically induced point spread function (PSF) from
short exposure images of two binary stars, HR 5138 and HR 5747 obtained at the
cassegrain focus of the 2.34 meter Vainu Bappu Telescope(VBT), situated at
Vainu Bappu Observatory(VBO), Kavalur. The position angles and separations of 
the binary components were seen to be consistent with results of the 
auto-correlation technique, while the Fourier phases of the reconstructed images 
were consistent with published observations of the binary orbits.
\vspace{0.3cm}

\vspace{0.3cm}

\begin{center} 
{\bf 1. Introduction} 
\end{center} 
\vspace{0.3cm}

\noindent
Atmospheric turbulence degrades the images obtained by ground based astronomical
telescopes. Schemes like speckle interferometry (Labeyrie, 1970), Knox-Thomson
algorithm (Knox and Thompson, 1974), shift and add (Lynds {\it et al.}, 1976),
triple correlation (Lohmann {\it et al.}, 1983) have been successfully
employed to restore the degraded images. All these schemes depend on the
statistical treatment of a large number of images. Often, it may not be possible
to record a large number of images within the time interval over which the
statistics of the atmospheric turbulence remains stationary. In such cases,
where only few images can be used, there are a number of schemes to restore the 
image using some prior information about the image. The maximum entropy method 
(Jaynes, 1982), CLEAN algorithm (Hogbom, 1974), and BID (Ayers and Dainty, 1988) 
are examples of such schemes. In this paper, we employ a version of BID 
developed by P. Nisenson (Nisenson, 1991), on degraded images of two binaries, 
HR 5138 HR 5138 (${m_v}$=5.57, $\Delta$m=0.2), and HR 5747 (${m_v}$=3.68, 
$\Delta$m=1.5) obtained at the 2.34 meter VBT at Kavalur. 
\vspace{0.3cm}

\begin{center} 
{\bf 2. Observations} 
\end{center}
\vspace{0.3cm}

\noindent
The 2.34 meter VBT has two accessible foci for backend instrumentation $-$ a 
prime focus (f/3.25 beam) and a cassegranian focus (f/13 beam). The latter was 
used for the observations described in this paper. The cassegranian focus has 
an image scale of 6.7 arcseconds per mm. This was further magnified to 
$\sim$1.21 arcseconds per mm, using a Barlow lens arrangement (Saha {\it et al.}, 
1987, Chinnappan {\it et al.},1991). This enlarged image was recorded through a 
5 nm filter centred on H$\alpha$ using an EEV intensified CCD camera which 
provides a standard CCIR video output of the recorded scene. The interface 
between the intensifier screen and the CCD chip is a fibre-optic bundle which 
reduces the image size by a factor of 1.69. A DT-2851 frame grabber card 
digitises the video signal. This digitiser resamples the pixels of each row 
(385 CCD columns to 512 digitized samples) and introduces a net reduction in the
row direction of a factor of 1.27. 
\vspace{0.2cm}

\noindent
The frame grabber can store upto two images on its onboard memory. These
images are then written onto the hard disc of a personal computer. The
images were stored on floppy diskettes and later analysed on a pentium.
The observing conditions were fair with an average seeing of $\sim$2 arcseconds
during the nights of 16/17 March 1990. 
\vspace{0.3cm}

\begin{center} 
{\bf 3. Data Processing} 
\end{center}
\vspace{0.3cm}

\noindent
The blind iterative deconvolution technique is described in detail in the
literature (Bates and McDonnell, 1986). Essentially, it consists of using
very limited information about the image, like positivity and image size,
to iteratively arrive at a deconvolved image of the object, starting from
a blind guess of either the object or both the convolving function. The
implementation of the particular version of BID used by us is described
in Nisenson (1991).
\vspace{0.2cm}

\noindent
The algorithm has the degraded image ${c(x,y)}$ as the operand.  An initial
estimate of the point spread function (PSF) ${p(x,y)}$ has to be provided.
The degraded image is deconvolved from the guess PSF by Wiener filtering,
which is an operation of multiplying a suitable Wiener filter (constructed
from the Fourier transform ${P(u,v)}$ of the PSF) with the Fourier transform 
${C(u,v)}$ of the degraded image as follows
\vspace{0.2cm}
 
\begin{math}
{O(u,v) = C(u,v) {\frac {P^{*}(u,v)} 
{P(u,v)P^{*}(u,v) + N(u,v)N^{*}(u,v)}}} 
\end{math} 
\vspace{0.2cm}

where $O$ is the Fourier transform of the De-convolved image and $N$ is the
noise spectrum. 
\vspace{0.2cm}

\noindent
This result $O$ is transformed to image space, the negatives in the image
are set to zero, and the positives outside a prescribed domain (called
object support) are set to zero. The average of negative intensities
within the support are subtracted from all pixels. The process is repeated
until the negative intensities decrease below the noise.
\vspace{0.2cm}

\noindent 
A new estimate of the psf is next obtained by Wiener filtering the original
image ${c(x,y)}$ with a filter constructed from the constrained object
${o(x,y)}$. This completes one iteration.  This entire process is repeated until 
the derived values of ${o(x,y)}$ and ${p(x,y)}$ converge to sensible solutions.
\vspace{0.3cm}

\begin{center} 
{\bf 4. Results} 
\end{center}
\vspace{0.3cm}

\noindent
The results for HR 5138 and HR 5747 were arrived at after 350 iterations. Since 
the intensity of the stars were different, the value of the Weiner filter 
parameters also had to be chosen accordingly. Figures 1(a), 1(b), 1(c) show the 
speckle image, PSF, and deconvolved image of HR 5138 respectively. Figures 2(a), 
2(b), 2(c) are for HR 5747 respectively. The companions of HR 5138 and HR 5747 are 
separated by {0.27} and {0.20} arcseconds respectively. Measurements of position 
angle (235 and 110 degrees for HR 5138 and HR 5747 respectively) and separation 
were compatible with the values published in the CHARA catalogue (McAlister and 
Hartkopf, 1988). Horch (1994), too found similar results. The magnitude 
difference for the reconstructed objects were 0.04 and 1.65 respectively for 
HR 5138 and HR 5747. Although these values compare quite well with those
published in the Bright Star Catalogue (Hoffleit and Jaschek, 1982)
one needs to treat many more objects before one can characterise the
photometric quality of the reconstructions.
\vspace{0.3cm}

\begin{center} 
{\bf 5. Discussion and Conclusions} 
\end{center} 
\vspace{0.3cm}

\noindent
The present scheme of BID has the chief problem of convergence. It is indeed an 
art to decide when to stop the iterations. The results are also vulnerable to 
the choice of various parameters like the support radius, the level of high 
frequency suppression during the Wiener filtering, etc. The availability of 
prior knowledge on the object through autocorrelation of the degraded image was 
found to be very useful for specifying the object support radius. In spite of 
this care taken in the choice of the support radius, the psf for each star 
contains residual signatures of the binary sources. Although suggestions have 
been made for improving the convergence (cf.\ Jefferies and Christou, 1993), 
these improved algorithms require more than a single speckle frame. 
\vspace{0.2cm}

\noindent
For the present, it is noteworthy that useful reconstructions are possible using 
single speckle frames. Questions regarding their dynamic range and linearity can 
be answered only after examining a wide range of reconstructions. The iterative 
nature of the algorithm does not lend to an explicit estimation of these 
parameters from a limited sample. New developments in camera electronics promise 
the capability to acquire several images within a short time. This will also 
reduce the level of artifacts. The chief achievement in this paper is a 
demonstration of the scientific potential of BID for resolving bright objects 
acquired using simple apparatus.
\vspace{0.5cm}

\noindent
\begin{figure}
\centerline{\psfig{figure=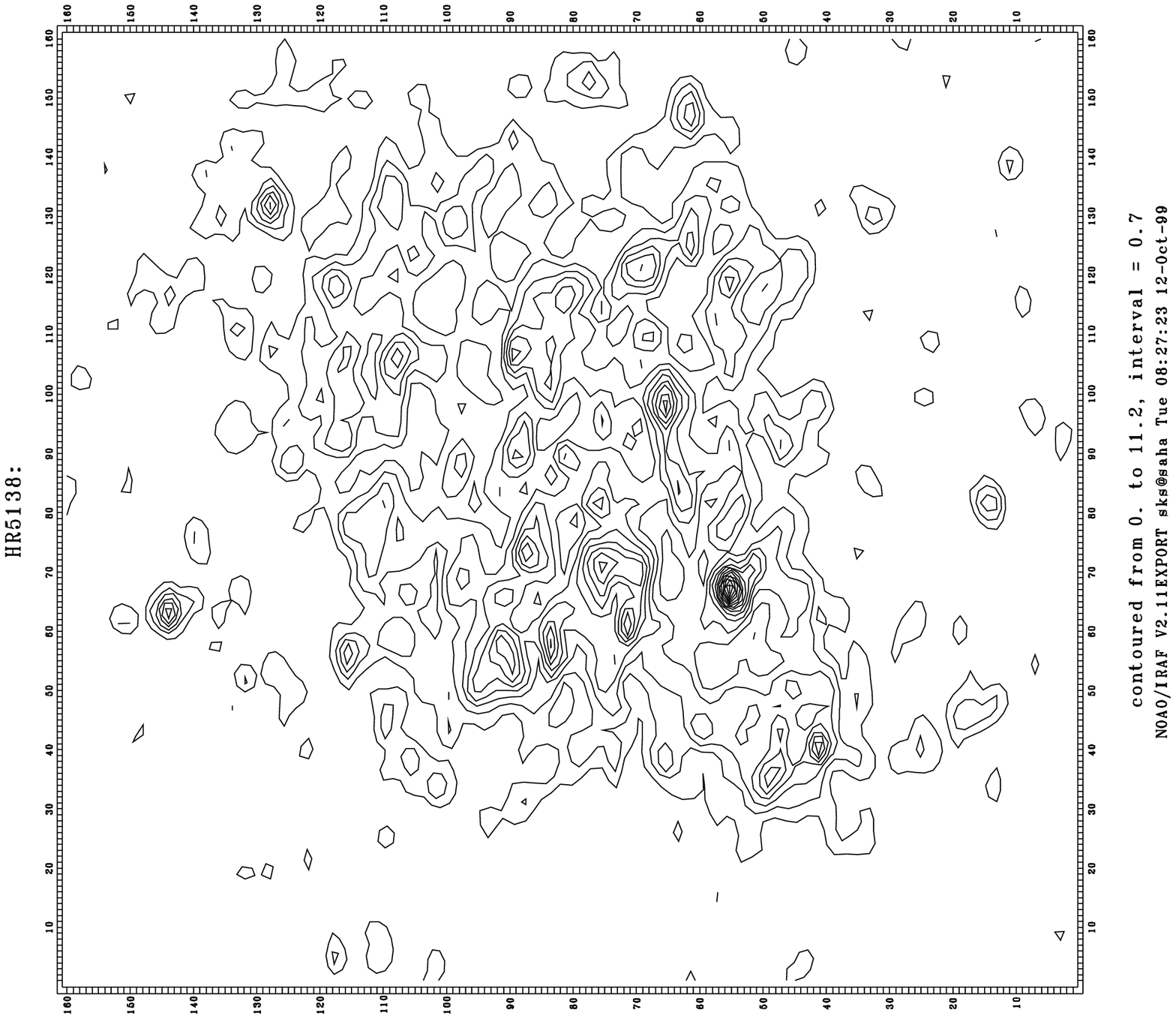,height=6cm,width=8cm,angle=270}}
%\caption{ }
\end{figure}

\noindent
\begin{figure}
\centerline{\psfig{figure=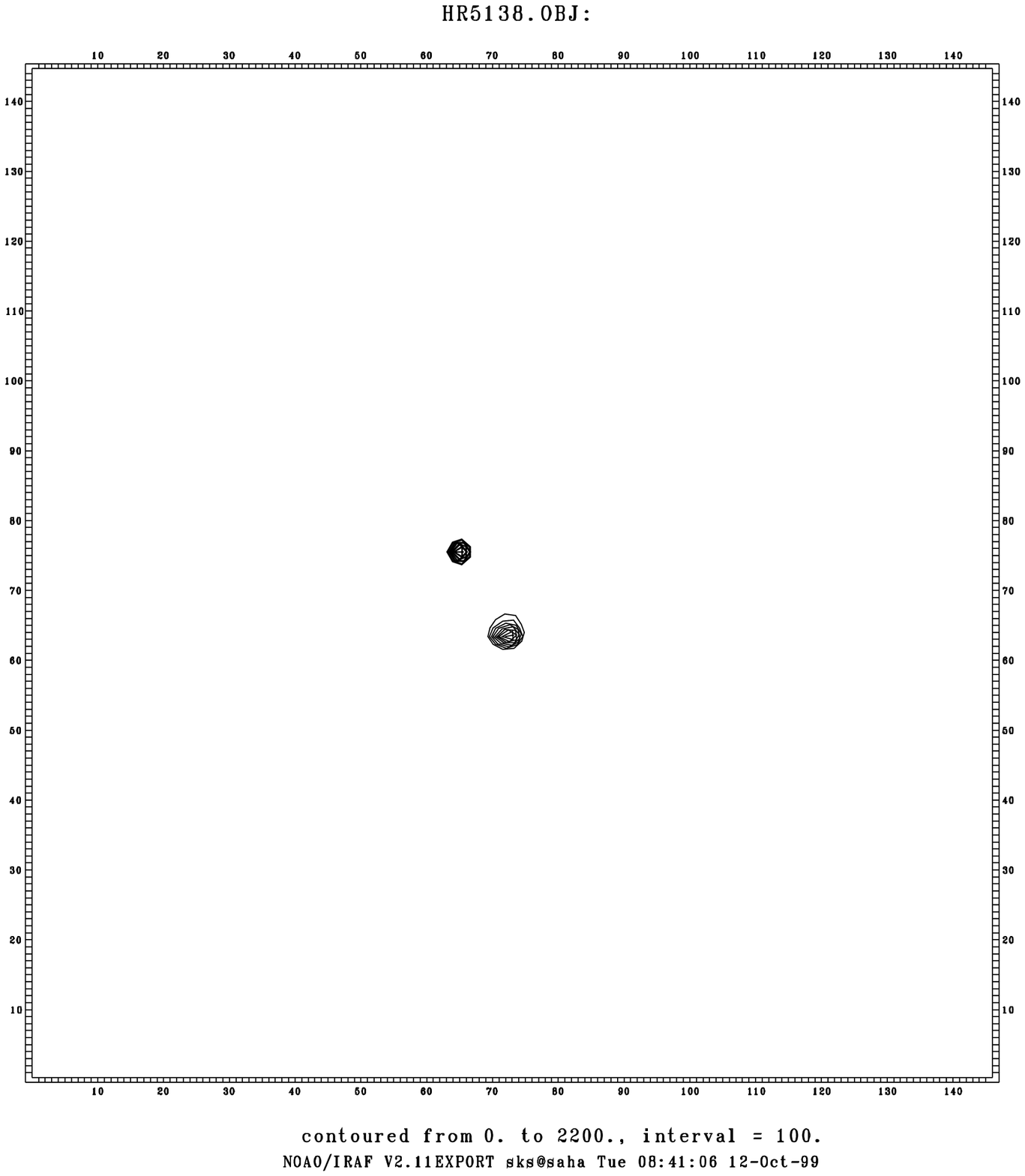,height=6cm,width=8cm,angle=270}}
%\caption{ }
\end{figure}

\noindent
\begin{figure}
\centerline{\psfig{figure=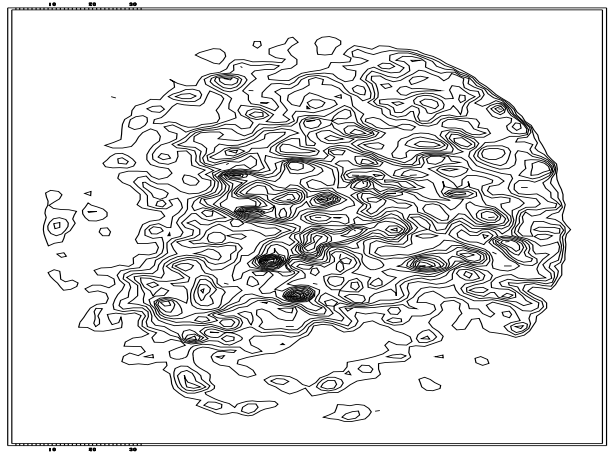,height=6cm,width=8cm,angle=270}}
%\caption{ }
\caption{1(a), 1(b), 1(c) show the speckle image, PSF, and deconvolved image 
of HR~5138 respectively. The numbers on the axes denote pixel numbers with
each pixel being equal to 0.02 arc sec.}
\end{figure}

\noindent
\begin{figure}
\centerline{\psfig{figure=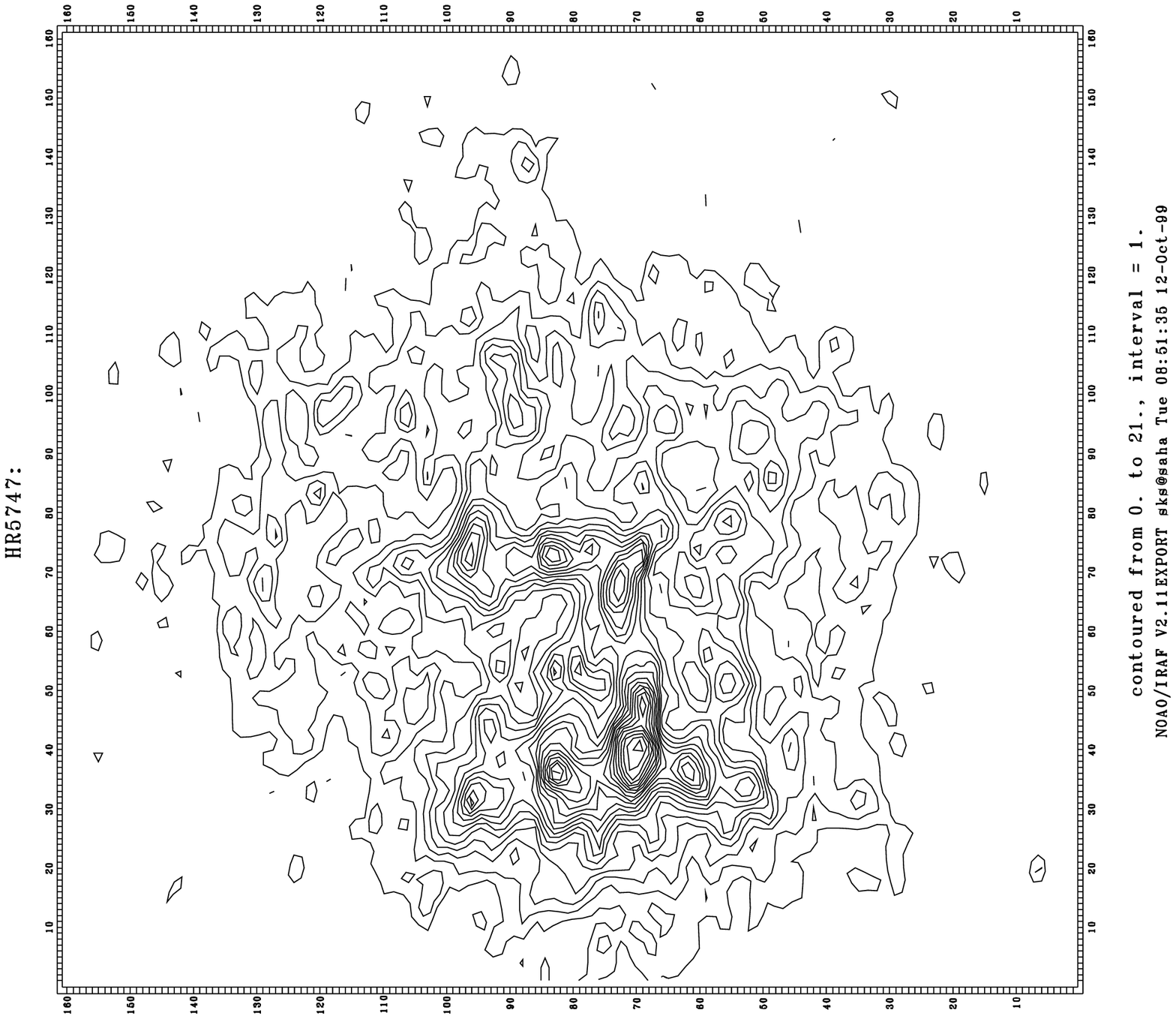,height=6cm,width=8cm,angle=270}}
%\caption{ }
\end{figure}

\noindent
\begin{figure}
\centerline{\psfig{figure=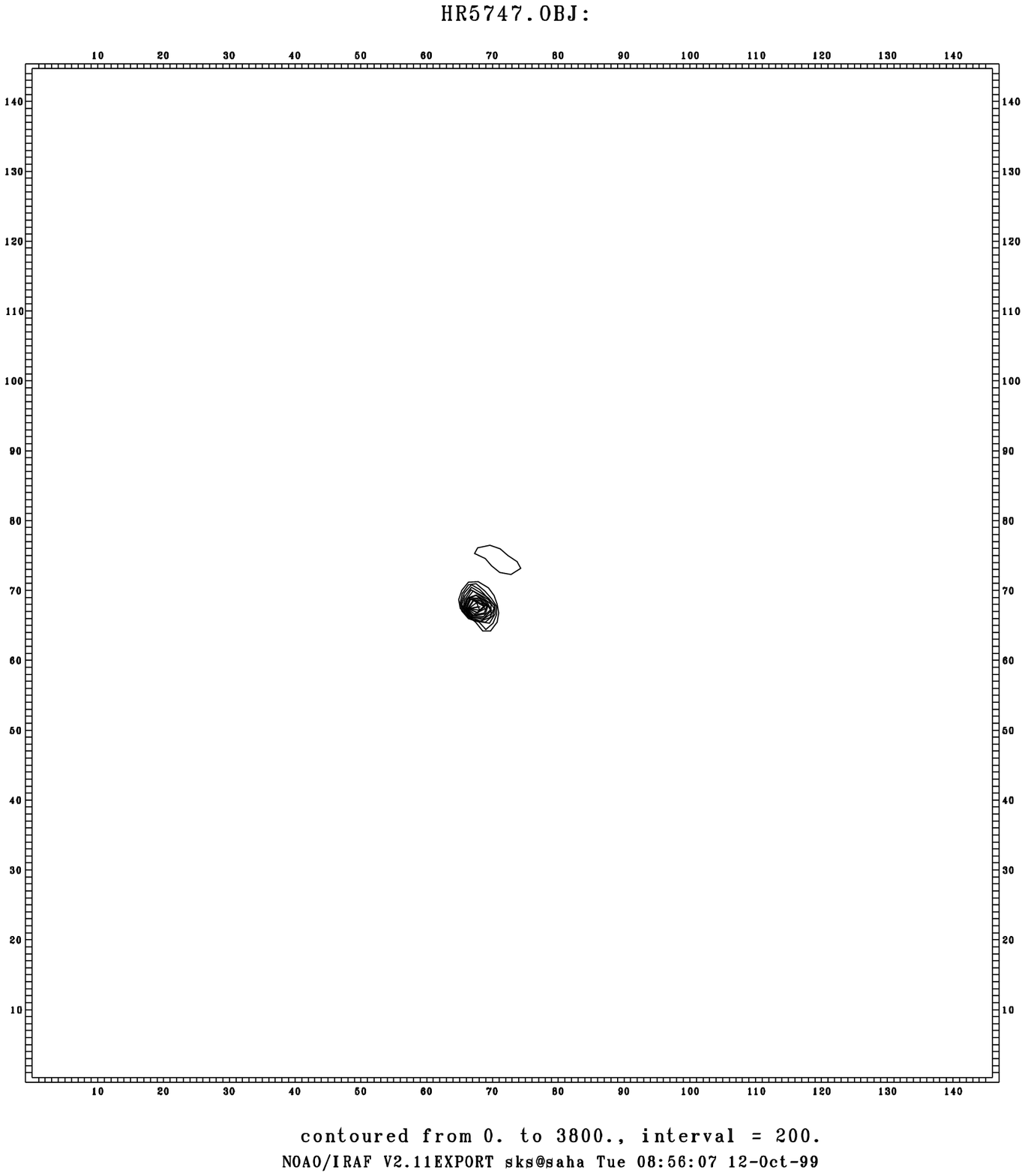,height=6cm,width=8cm,angle=270}}
%\caption{ }
\end{figure}

\noindent
\begin{figure}
\centerline{\psfig{figure=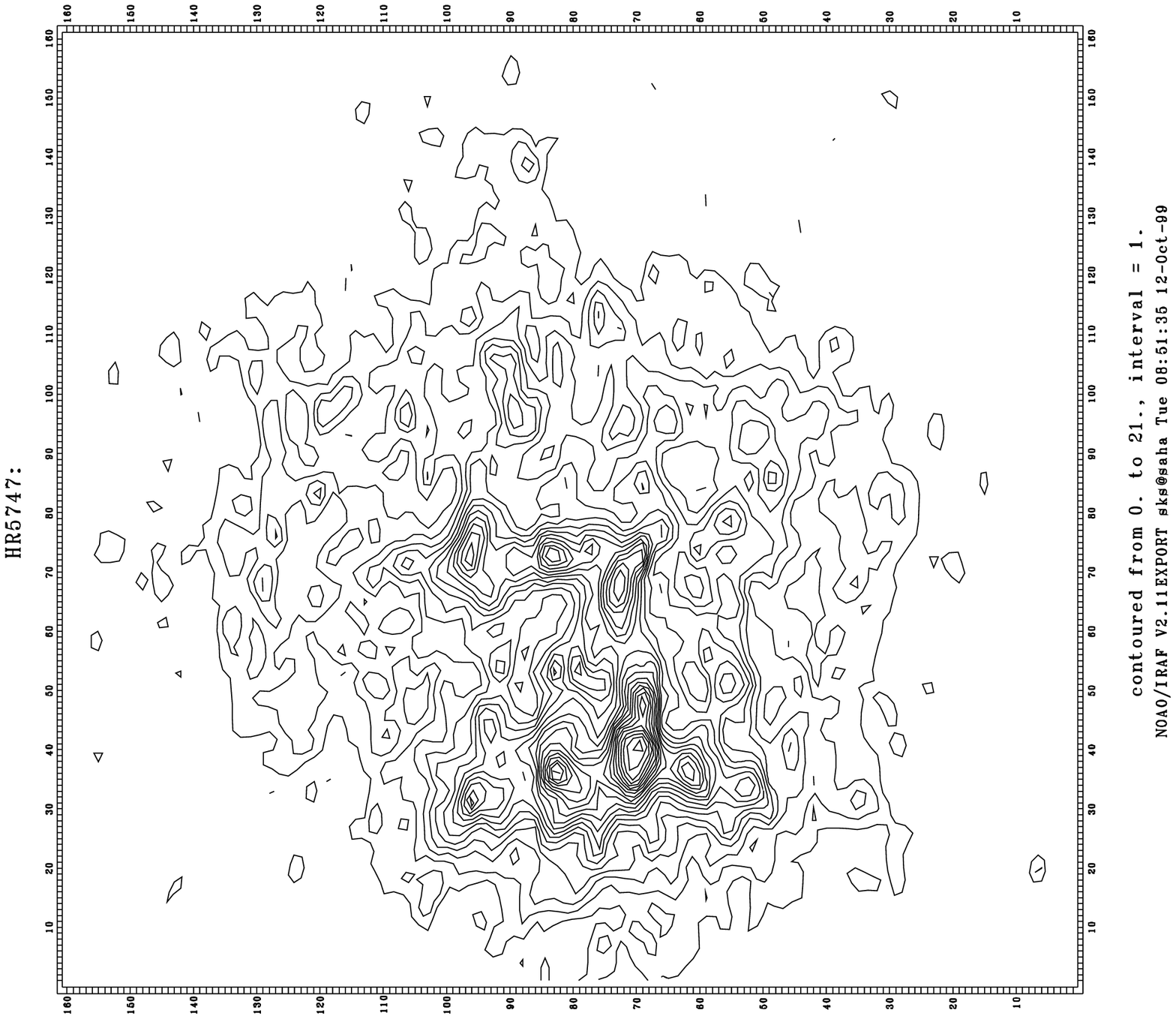,height=6cm,width=8cm,angle=270}}
%\caption{ }
\caption{2(a), 2(b), 2(c) are for HR~5747 respectively.} 
\end{figure}

\noindent
{\bf Acknowledgment} The authors are grateful to Dr. P. Nisenson of Center for 
Astrophysics, Cambridge, USA, for the BID code as well as for useful discussions.

\begin{center} 
{\bf References} 
\end{center} 
\vspace{0.3cm}

\noindent
Ayers G.R. \& Dainty J.C., 1988, Optics Letters, {\bf 13}, 547. \\
\noindent
Bates R.H.T. \& McDonnell M.J., 1986, "Image Restoration and Reconstruction",
Oxford Engineering Science {\bf 16}, Clarendon Press, Oxford. \\
\noindent
Chinnappan V., Saha S.K. \& Faseehana, 1991, Kod. Obs. Bull. {\bf 11}, 87. \\
\noindent
Hoffleit D. \& Jaschek, C., 1982, "The Bright Star Catalogue",
Yale University Observatory \\
\noindent
Hogbom J., 1974, Ap.J. Suppl., {\bf 15}, 417. \\
\noindent
Horch E. P., 1994, Ph.D Thesis, Stanford University \\ 
\noindent
Jaynes E.T., 1982, Proc. IEEE, {\bf 70}, 939. \\
\noindent
Jefferies S.M. \& Christou J.C., 1993, ApJ., {\bf 415}, 862. \\
\noindent
Knox K.T. \& Thompson B.J., 1974, Ap.J. Lett., {\bf 193}, L45. \\ 
\noindent
Labeyrie A., 1970, A and A, {\bf 6}, 85. \\
\noindent
Lohmann A.W., Weigelt G., Wirnitzer B., 1983, Appl. Opt., {\bf 22}, 4028. \\ 
\noindent
Lynds C.R., Worden S.P., Harvey J.W., 1976, Ap.J., {\bf 207}, 174. \\
\noindent
McAlister H. A. \& Hartkopf W. I., 1988, CHARA contribution no. 2. \\
\noindent
Nisenson P., 1991, in Proc. ESO-NOAO conf. on High Resolution Imaging
by Interferometry ed. J M Beckers \& F Merkle, p-299. \\
\noindent
Saha S.K., Venkatakrishnan P., Jayarajan A.P. \& Jayavel N., 1987, Current
Science, {\bf 56}, 985. \\

\end{document}